\documentclass[aps, prl, twocolumn, superscriptaddress, showpacs]{revtex4}
\usepackage[ansinew]{inputenc}
\usepackage{epsfig}
\usepackage{graphicx}
\usepackage{color}
\usepackage{bm}
\usepackage{multirow}
\usepackage{mathrsfs}
\usepackage{revsymb}
\usepackage{amssymb}
\usepackage{amsmath}
\usepackage{slashed}
\usepackage{longtable}
\usepackage{dcolumn}
\usepackage{float}
\newcount\colveccount
\newcommand*\colvec[1]{
        \global\colveccount#1
        \begin{pmatrix}
        \colvecnext
}
\def\colvecnext#1{
        #1
        \global\advance\colveccount-1
        \ifnum\colveccount>0
                \\
                \expandafter\colvecnext
        \else
                \end{pmatrix}
        \fi
}

\begin{document}

\title{Non-perturbative analysis of nuclear shape effects on the bound electron g factor}

\author{Niklas Michel}
\email[]{niklas.michel@mpi-hd.mpg.de}
\affiliation{Max Planck Institute for Nuclear Physics, Saupfercheckweg~1, 69117 Heidelberg, Germany}

\author{Jacek Zatorski}
\affiliation{Max Planck Institute for Nuclear Physics, Saupfercheckweg~1, 69117 Heidelberg, Germany}

\author{Natalia S. Oreshkina}
\affiliation{Max Planck Institute for Nuclear Physics, Saupfercheckweg~1, 69117 Heidelberg, Germany}

\author{Christoph H. Keitel}
\affiliation{Max Planck Institute for Nuclear Physics, Saupfercheckweg~1, 69117 Heidelberg, Germany}

\date{\today}

\begin{abstract}
The theory of the $g$ factor of an electron bound to a deformed nucleus is considered non-perturbatively and results are presented for a wide range of nuclei with charge numbers from ${Z}{=}{16}$ up to ${Z}{=}{98}$. We calculate the nuclear deformation correction to the bound electron $g$ factor within a numerical approach and reveal a sizable difference compared to previous state-of-the-art analytical calculations. We also note particularly low values in the region of filled proton or neutron shells, and thus a reflection of the nuclear shell structure both in the charge and neutron number.\\
\end{abstract}


\maketitle

The electron's $g$ factor characterizes its magnetic moment in terms of its angular momentum. For an electron bound to an atomic nucleus, the $g$ factor can be predicted in the framework of bound state quantum electrodynamics (QED) as well as measured in Penning traps, both with a very high degree of accuracy. This enables extraction of information on fundamental interactions, constants and nuclear structure. For example, the combination of theory and precise measurements of the bound electron $g$ factor has recently provided an enhanced value for the electron mass~\cite{Sturm2014}, and bound state QED in strong fields was tested with unprecedented precision~\cite{Haffner2000, Verdu2004, Kohler2015, Zatorski2017}. It also enables measurement on characteristics of nuclei such as electric charge radii, as shown for $\textrm{Si}^{13+}$ ion~\cite{Sturm2011}, or the isotopic mass difference as demonstrated for $\null^{48} \textrm{Ca}$ and $\null^{40} \textrm{Ca}$ in~\cite{Kohler2016}, or, as proposed theoretically, magnetic moments~\cite{Yerokhin2011}.  Also, it was argued that $g$-factor experiments with heavy ions could result in a value for the fine-structure constant which is more accurate than the presently established one~\cite{Shabaev2006}.
With planned experiments involving high $Z$ nuclei~\cite{HITRAP2008,vogel2015} and current experimental accuracies on the $10^{-10}$ level for low $Z$, it is important to keep track also of higher order effects. In this context, the influence of nuclear size and deformation is critical. In~\cite{jacek2012}, the nuclear shape correction to the bound electron $g$ factor was introduced and calculated for spinless nuclei using the perturbative effective radius method~\cite{Shabaev1993,kozhedub2008}. This effect takes the influence of a deformed nuclear charge distribution into account, and changes the g factor on a $10^{-6}$ level for heavy nuclei, thus being potentially visible in future experiments. Therefore, a comparison of experiment and theory for heavy nuclei demands a critical scrutiny of the validity of the previously used perturbative methods, as pointed out in \cite{karshenboim2018}.

In this paper, we present non-perturbative calculations of the nuclear deformation correction to the bound electron $g$ factor and show the corresponding values for nuclei across the entire nuclear chart, quantifying the non-perturbative corrections and especially observing the appearance of nuclear shell closure effects in the values of the bound electron $g$ factor.

Relativistic units with ${\hbar}{=}{c}{=}{1}$ are used throughout this work, as well as the Heavyside unit of charge with ${\alpha}{=}{e^2}/{4\pi}$, where $\alpha$ is the fine structure constant and the elementary charge $e$ is negative.

It has been shown in \cite{Kozhedub} that for spinless nuclei the relativistic Hamiltonian for the electron bound to a deformed nucleus reads
\begin{equation}
\text{H}_{e}=\vec{\alpha}\cdot\vec{p} + \beta m_e + V(r).
\label{eq:Hboundel}
\end{equation}
Here, $\vec{\alpha}$ and $\beta$ are the four Dirac matrices, $\vec{p}$ is the electron's momentum, $m_e$ the electron mass, and the electric interaction between eletron and nucleus can be described in terms of the nuclear charge distribution $\rho(\vec{r})$ as
\begin{equation}
V(r) = -Z \alpha \int \text{d}^3r^{'}\,\frac{\rho(\vec{r}^{\,'})}{r_>},
\label{eq:potdef}
\end{equation}
where ${r_>}{:=}{\text{max}(r,r^{'})}$. For spherically symmetric charge distributions, this leads to finite size effects in atomic spectra \cite{Shabaev1993}. However it is important to note, that this formula is also valid for deformed nuclear charge distributions, although the resulting potential is spherically symmetric. The solution of the corresponding eigenvalue equation
\begin{equation}
\text{H}_{e} \left| n\kappa m \right> = E \left| n \kappa m \right>
\label{eq:diraceq}
\end{equation}
can be written in position space in terms of the well-know spherical spinors $\Omega_{\kappa m}(\vartheta,\varphi)$ and the radial functions $G_{n\kappa}(r)$, $F_{n\kappa}(r)$~\cite{greiner2000}, and depends on the principal quantum number $n$, the relativistic angular momentum quantum number $\kappa$, and the $z$-component of the total angular momentum $m$.

In this work, we focus on quadrupole deformations, since atomic nuclei do not possess static dipole moments. Here, the deformed Fermi distribution
\begin{equation}
\rho_{ca\beta}(r,\vartheta)=\cfrac{N}{1+\text{exp}(\frac{r-c(1+\beta \text{Y}_{20}(\vartheta))}{a})}
\label{eq:deffermi}
\end{equation}
as a model of the nuclear charge distribution has proved to be very successful, e.g. in heavy muonic atom spectroscopy with deformed nuclei \cite{hitlin1970,tanaka1984}; the normal Fermi distribution (${\beta}{=}{0}$) has also been used in electron-nucleus scattering experiments determining the nuclear charge distribution \cite{hahn1956}. Here, $a$ is a skin thickness parameter and $c$ the half-density radius, while $\beta$ is a deformation parameter. $\text{Y}_{lm}(\vartheta,\varphi)$ are the spherical harmonics and $\text{Y}_{l0}(\vartheta)$ depend only on the polar angle $\vartheta$, and not on the azimuthal angle $\varphi$. The normalization constant $N$ is determined by the condition
\begin{equation}
\int \text{d}^3r\, \rho_{ca\beta}(r,\vartheta)=1.
\end{equation}

In an external, homogeneous, and weak magnetic field~$\vec{B}$, the $g$~factor of the bound electron is defined by the first order energy splitting $\delta E$ due to the external field as the proportionality coefficient \cite{Beier2000}
\begin{equation}
\delta E=-e \,\left< n\kappa m \right|\vec{\alpha}\cdot \vec{A}\left|n\kappa m\right> =:m \, g\,  \mu_B {|}{\vec{B}}{|},
\end{equation}
where $\vec{A}=\frac{1}{2}[\vec{B}\times\vec{r}\,]$ is the corresponding vector potential, and $\mu_B$ the Bohr magneton.
The $g$ factor in central potentials is independent of the quantum number $m$. It can be expressed in terms of the radial functions as
\begin{equation}
g=\frac{2m_e \kappa}{j(j+1)} \int_0^\infty \text{d}r\, r\, G_{n \kappa}(r)\,F_{n\kappa}(r),
\label{eq:gfacgs}
\end{equation}
where $j=\left|\kappa\right|-1/2$ is the total angular momentum of the electron. Alternatively, the bound electron $g$ factor can be expressed for arbitrary central potentials in terms of the derivative of the eigenenergies with respect to the electron's mass \cite{Karshenboim2005} as
\begin{equation}
g=-\frac{\kappa}{2j(j+1)}\left( 1-2\kappa\frac{\partial E_{n\kappa}}{\partial m_e} \right).
\label{eq:deriv}
\end{equation}

\begin{table}[b]
\caption{\label{tab:spline}%
Comparison of the nuclear deformation $g$ factor correction obtained by the effective radius method ($\delta g_{\text{ND}}^{(\text{eff})}$) and non-perturbatively by numerical calculations ($\delta g_{\text{ND}}^{(\text{num})}$) for several isotopes. $R_N$ is the RMS nuclear electric charge radius from literature \citep{Angeli2013} and $\beta$ is the parameter of the deformed Fermi distribution (\ref{eq:deffermi}). Both methods, as well as the procedure for obtaining the parameters of the deformed Fermi distribution, are described in the text.
}
\begin{ruledtabular}
\begin{tabular}{lcccc}
Isotope & $R_N(\text{fm})$  & $\beta$ & $\delta g_{\text{ND}}^{(\text{eff})}$ & $\delta g_{\text{ND}}^{(\text{num})}$\\
\colrule\\[-5pt]
$^{\phantom{0}58}_{\phantom{0}26}$Fe & 3.775 & 0.273 & $-2.19\times 10^{-11}$ & $-2.11\times 10^{-11}$\\[4pt]
$^{\phantom{0}82}_{\phantom{0}38}$Sr & 4.246 & 0.263 & $-3.56\times 10^{-10}$ & $-3.27\times 10^{-10}$\\[4pt]
$^{\phantom{0}98}_{\phantom{0}44}$Ru & 4.423 & 0.194 & $-6.00\times 10^{-10}$ & $-5.41\times 10^{-10}$\\[4pt]
$^{116}_{\phantom{0}48}$Cd  &  4.628  & 0.189 & $-1.20\times 10^{-9\phantom{0}}$ & $-1.07\times 10^{-9\phantom{0}}$ \\[4pt]
$^{116}_{\phantom{0}50}$Sn  &  4.627  & 0.108 & $-5.11\times 10^{-10}$ & $-4.55\times 10^{-10}$ \\[4pt]
$^{134}_{\phantom{0}54}$Xe  &  4.792  & 0.113 & $-1.09 \times 10^{-9\phantom{0}}$ & $-9.62 \times 10^{-10}$ \\[4pt]
$^{152}_{\phantom{0}64}$Gd  &  5.082  & 0.202 & $-1.53\times 10^{-8\phantom{0}}$ & $-1.32\times 10^{-8\phantom{0}}$ \\[4pt]
$^{208}_{\phantom{0}82}$Pb  &  5.501  & 0.061 & $-1.35\times 10^{-8\phantom{0}}$ & $-1.13\times 10^{-8\phantom{0}}$ \\[4pt]
$^{244}_{\phantom{0}94}$Pu  &  5.864  & 0.287 & $-1.28\times 10^{-6\phantom{0}}$ & $-1.05\times 10^{-6\phantom{0}}$ \\[4pt]
$^{248}_{\phantom{0}96}$Cm  & 5.825 & 0.299 & $-1.70\times 10^{-6\phantom{0}}$ & $-1.39\times 10^{-6\phantom{0}}$\\[4pt]
\end{tabular}
\end{ruledtabular}
\end{table}

For the model of a point-like nucleus, the radial wave functions are known analytically and the ground state $g$ factor with $n=1$ and $\kappa=-1$ is
\begin{equation}
g_{\text{point}} = \frac{2}{3}\left( 1 + 2 \gamma\right),
\end{equation}
with $\gamma = \sqrt{1-(Z\alpha)^2}$, a result presented for the first time by Breit$\,$\cite{breit1928}. For the deformed Fermi distribution~(\ref{eq:deffermi}) with a fixed charge number $Z$, the $g$ factor (\ref{eq:gfacgs}) is completely determined by the parameters $c$, $a$ and $\beta$, and therefore can be written for the ground state as
\begin{equation}
g = g_{\text{point}} + \delta g^{(ca\beta)}_{\text{FS}},
\end{equation}
where $\delta g^{(ca\beta)}_{\text{FS}}$ is the finite size correction depending on the parameters $c$, $a$, and $\beta$. In \cite{jacek2012}, the nuclear deformation correction to the bound electron $g$ factor is defined as the difference of the finite size effect due a deformed charge distribution and due to a symmetric charge distribution (i.e. ${\beta}{=}{0}$) with the same nuclear radius as
\begin{equation}
\delta g_{\text{ND}}=\delta g^{(c_1a\beta)}_{\text{FS}} - \delta g^{(c_2a0)}_{\text{FS}},
\label{eq:defdgnd}
\end{equation}
where $a=2.3\,\text{fm}/(4\text{ln}(3))$, and $c_i$ are determined such that $\sqrt{\left<r^2\right>_{\rho}}$ of the corresponding charge distribution agrees with the root-mean-square (RMS) values from literature \cite{Angeli2013}. The $n$-th moment of a charge distribution $\rho(\vec{r}\,)$ is defined as
\begin{equation}
\left< r^n \right>_{\rho} = \int \text{d}^3r\,\, r^n \rho(\vec{r}\,).
\label{eq:nmoment}
\end{equation}
Values for the deformation parameter $\beta$ can be obtained by literature values of the reduced E2-transition probabilities from a nuclear state $2^+_i$ to the ground state $0^+$ via~\cite{Trager}:
\begin{equation}
\beta = \frac{4\pi}{3Z|e|\sqrt{5\left< r^2\right>_{\rho} /3}}\left[ \sum_i B(E2;0^+\rightarrow 2_i^+) \right]^{1/2}
\label{eq:beta}
\end{equation}
From Eq. (\ref{eq:defdgnd}) it is evident that the nuclear deformation correction is a difference of two finite size effects and therefore especially sensitive on higher order effects. However, for high $Z$ it reaches the $10^{-6}$~level and therefore is very significant.

It was shown in~\cite{jacek2012} with the effective radius method \cite{Shabaev1993} that $\delta g_{\text{FS}}^{(ca\beta)}$ and therefore $\delta g_{\text{ND}}$ mainly depends on the moments $\left< r^2 \right>_{\rho}$ and $\left< r^4 \right>_{\rho}$. $\delta g_{\text{ND}}$ can be calculated with the formula \cite{Karshenboim2005}
\begin{equation}
\delta g^{(ca\beta)}_{\text{FS}}=\frac{4}{3}\frac{\partial E_{\text{FS}}}{\partial m_e},
\label{eq:effrad}
\end{equation}
which is a direct consequence of Eq. (\ref{eq:deriv}).
The energy correction due to $\rho_{ca\beta}(r,\vartheta)$ compared to the point like nucleus can be approximated as \cite{Shabaev1993}
\begin{equation}
E_{\text{FS}}\approx\frac{(Z\alpha)^2}{10}\left[{1}{+}(Z\alpha)^2f(Z\alpha) \right](2Z\alpha R m_e)^{2\gamma}m_e.
\label{eq:efs}
\end{equation}
Here, $f(x)=1.380-0.162x+1.612x^2$ and the radius of a homogeneously charged sphere with approximately the same energy correction as the original charge distribution is
\small
\begin{equation}
R=\sqrt{\frac{5}{3}\left<r^2\right>_{\rho_{ca\beta}}\left[ 1-\frac{3}{4}(Z\alpha)^2 \left( \frac{3}{25}\cfrac{\left<r^4\right>_{\rho_{ca\beta}}}{\left<r^2\right>^2_{\rho_{ca\beta}}}-\frac{1}{7} \right) \right]}.
\label{eq:radius}
\end{equation}
\normalsize
While \eqref{eq:effrad} is exact for an arbitrary central potential, provided that $E_{\text{FS}}$ is known exactly, \eqref{eq:efs} is an approximation derived under the assumption of the difference between point-like and extended potential being a perturbation. The calculation of the nuclear deformation correction to the bound electron $g$ factor via the effective radius approach therefore relies on a perturbative evaluation of the energy derivative in Eq.~(\ref{eq:effrad}) and is limited by the accuracy of the finite size corrections.

In this work, we also calculate $\delta g_{\text{ND}}$ non-perturbatively by solving the Dirac equation (\ref{eq:diraceq}) numerically for the potential (\ref{eq:potdef}), including all finite size effects due to the deformed charge distribution $\rho_{ca\beta}(r,\vartheta)$. The dual kinetic balance method \cite{dualkinetic} is used for numerical calculations. Then, the $g$ factors needed in Eq.~(\ref{eq:defdgnd}) for the nuclear deformation correction can be obtained by numerical integration of the wave functions in Eq. (\ref{eq:gfacgs}). Alternatively, the derivative of the energies in Eq. (\ref{eq:deriv}) can be approximated numerically as
\begin{equation}
\frac{\partial E_{n\kappa}}{\partial m_e} \approx \frac{E_{n\kappa}^{(m_e+\delta m)}-E_{n\kappa}^{(m_e-\delta m)}}{2 \delta m},
\end{equation}
with a suitable ${\delta m / m_e}{\ll}{1}$. Here, $E_{n\kappa}^{(m_i)}$ stands for the binding energy obtained by solving the Dirac equation with the electron mass replaced by $m_i$. We find both methods in excellent agreement.


\begin{figure*}
  \centering
  \begin{minipage}[b]{\textwidth}
    \includegraphics[width=0.68\textwidth]{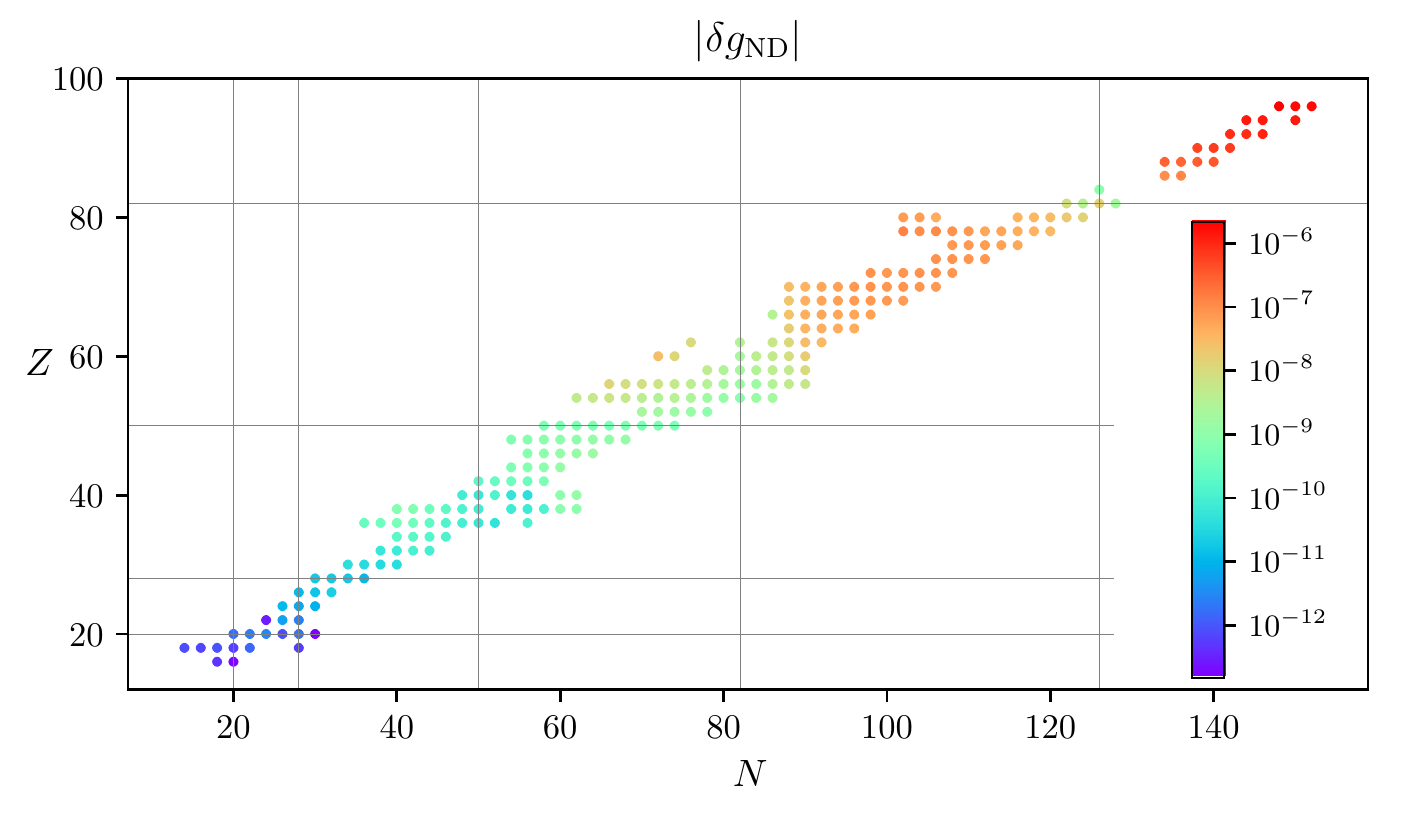}\\
    $\qquad \, \,$(a)
  \end{minipage}
  \hfill
  \begin{minipage}[b]{\textwidth}
    \includegraphics[width=0.99\textwidth]{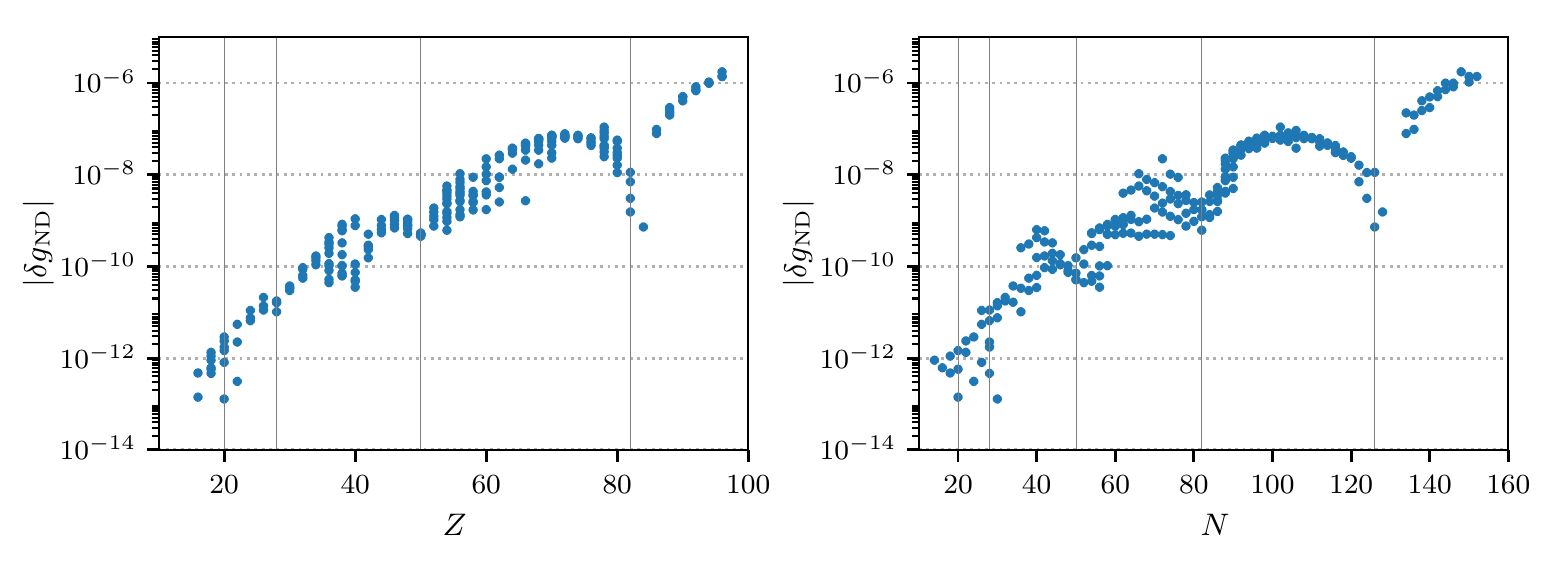}\\
    \hspace{1.3cm}(b)\hspace{8.3cm}(c)
    \caption{\label{fig:dg}(Color online) Nuclear chart with charge number $Z$ and neutron number $N$, where the grey lines indicate the magic numbers 20, 28, 50, 82, and 126. The points represent even-even nuclei, where their color in (a) displays the nuclear deformation g-factor correction $\delta g_{\text{ND}}$, which takes particularly low values around the magic numbers and larger values in between. The two lower figures show $\delta g_{\text{ND}}$ for the considered even-even nuclei as a function of only the charge number $Z$ (b) and of only the neutron number $N$ (c), respectively. The vertical solid grey lines are the nuclear magic numbers, which show that filled proton, as well as neutron shells, reduce $\delta g_{\text{ND}}$.}
  \end{minipage}
\end{figure*}

We calculated the nuclear deformation $g$-factor correction for a wide range of even-even, both in the proton and neutron number, and therefore spinless nuclei with charge numbers between 16 and 96 using the deformed Fermi distribution from Eq. (\ref{eq:deffermi}) with parameters $a$, $c$, and $\beta$ obtained as described below Eq.~(\ref{eq:defdgnd}). The numerical approach as described above was used, which does not rely on an expansion in $Z\alpha$ or in moments of the nuclear charge distribution. In Table \ref{tab:spline}, a comparison between this numerical approach and the effective radius method from \cite{jacek2012} shows that the latter is a good order-of-magnitude estimate of the nuclear deformation correction,
but for high-precision calculations,
non-perturbative methods are indispensable. Eq.~\eqref{eq:efs} has an estimated relative uncertainty ${\scriptstyle\lesssim}\,0.2\,\%$~\cite{Shabaev1993} and there are several aspects, which limit the accuracy of the effective radius method: Firstly, Eq.~\eqref{eq:radius} uses only the second and fourth moment of the nuclear charge distribution for finding the effective radius of a charged sphere with the same energy levels, and neglects higher orders of the charge distribution. Secondly, the energy levels of the charged sphere are calculated by the approximate formula~\eqref{eq:efs}. Additionally, it was shown in \cite{karshenboim2018} that the effective radius method for arbitrary charge distributions is incomplete in order $(Z\alpha)^2m_e(Z\alpha m_e R_N)^3$, where $R_N$ is the nuclear RMS charge radius. Our numerical approach overcomes this issue by dispensing with an expansion both in $Z\alpha$ and $Z\alpha\, m_e\, R_N$.
Finally, the uncertainty in the energy correction might not transfer trivially to the uncertainty in the energy derivative with respect to the electron mass.
Being a difference of two small finite-nuclear-size corrections itself, the nuclear deformation correction can exhibit enhanced sensitivity on these factors. Therefore, the uncertainty of the finite-nuclear-size corrections via the effective radius method can lead to a sizable uncertainty for the nuclear deformation corrections, especially for high $Z$. Convergence of the numerical method, on the other hand, was checked by varying numerical parameters and using various grids, and the obtained accuracy permits the consideration of nuclear size and shape with an accuracy level much higher than the differences to the perturbative method.

The required RMS values for the nuclear charge radius are taken from \cite{Angeli2013} and the reduced transition probabilities needed for the calculation of $\beta$ via (\ref{eq:beta}) from \cite{ENSDF}. The resulting values for $|\delta g_{\text{ND}}|$ are shown in Fig. \ref{fig:dg} as a function of the charge number $Z$ and the neutron number $N$. If proton or neutron number is in the proximity of a nuclear magic number 2, 8, 20, 28, 50, 82, and 126, which corresponds to a filled proton or neutron shell \cite{Ring}, the nuclear shell closure effects also transfer to the bound electron $g$~factor, and the nuclear deformation correction is reduced.\\


Concluding, the nuclear deformation $g$-factor correction was calculated non-perturbatively for a wide range of nuclei by using quadrupole deformations estimated from nuclear data.
By comparing the previously used perturbative effective radius method and the all-order numerical approach, it was shown that the contributions of the non-perturbative effects can amount up to the $20\,\%$ level. 
In the low-$Z$ regime, the nuclear deformation corrections can safely be neglected, especially for the ions considered in \cite{Sturm2014}. However, considering a parts-per-million nuclear deformation correction and an expected parts-per-billion accuracy, or even below, for the g factor experiments with high-$Z$ nuclei, in this case an all-order treatment is indispensible.
These results motivate a non-perturbative treatment of nuclear shape effects also for other atomic properties, e.g. fine and hyperfine splittings.
The distribution of electric charge inside the nucleus is a major theoretical uncertainty for g factors with heavy nuclei, which suggests the extraction of information thereon from experiments. Our work demonstrates the required accurate mapping of arbitrary nuclear charge distributions to corresponding g factors.
Furthermore, the nuclear deformation correction was shown to be a not monotonically increasing function of the nuclear charge number. In fact, it rather reflects the nuclear shell model by taking particularly low values around filled proton, as well as neutron shells, showing the neutrons' indirect influence on the distribution of electric charge inside the atomic nucleus.

The authors acknowledge insightful conversations with Z. Harman and V. Debierre.


\end{document}